\documentclass[prl,nofootinbib,
noshowkeys,twocolumn,superscriptaddress]{revtex4}
\usepackage{graphicx,amsfonts}

\newcommand{\be}{\begin{eqnarray}}
\newcommand{\ee}{\end{eqnarray}}
\newcommand{\la}{\langle}
\newcommand{\ra}{\rangle}
\newcommand{\ta}{\overline{\theta}}

\begin{document}
\author{Pietro Faccioli}
\email{faccioli@ect.it}
\affiliation{Dipartimento di Fisica, Universit\'a degli Studi di Trento, 
Via Sommarive 15, 
Povo (Trento); E.C.T.*, Strada delle Tabarelle 286, 
I-38050 Villazzano (Trento); I.N.F.N. Sezione Collegata di Trento}
\title{Strong CP breaking and quark-antiquark repulsion in QCD,
at finite $\theta$}
\vspace {2cm}

\begin{abstract}
This work is devoted to the study of 
the CP-breaking dynamics in QCD, at finite $\theta$-angle.
By working in the semi-classical limit, in which the topology 
of the vacuum is clustered around instantons and anti-instantons,
we show that the quantum fluctuations of the $\theta$-vacuum
generate an effective flavor-dependent repulsion between matter and 
anti-matter. 
As a consequence, during the tunneling between the degenerate classical vacua,
quarks and anti-quarks in a neutron
migrate in opposite directions, giving rise to an oscillating 
electric dipole moment.
We discuss a possible phenomenological implication of this effect. 

\end{abstract}
\maketitle
The hypothesis of CP-violation in the strong sector
of the Standard Model arises from the analysis of the 
quantum structure of the QCD vacuum.
At a classical level, the QCD Hamiltonian
possesses an infinite set of degenerate gauge-dependent vacua, 
each of which is characterized by its winding number $n$. 
The gauge-invariant definition of the ground-state is 
given in terms of a linear superposition of such 
classical vacua~ ($\theta$-vacuum)~\cite{callan}:
\be
\label{thetavacuum}
|\theta\ra = \sum_{n} e^{i\,\theta\, n} |n\ra,\qquad \theta\in \mathbb{R}.
\ee

The structure of the QCD vacuum, combined with the observation  
of weak CP-violation in neutral kaon systems, gives rise to  
the so-called  $\theta$-term in the QCD action (in                  
the Euclidean formulation): 
\be
\label{Ltheta}
\mathcal{S}\to \mathcal{S}+S_{\theta},\qquad
\mathcal{S}_\theta= i\,\overline{\theta}
 \,\frac 1{32\pi ^2}\int d^4z\,F_{\mu \nu }%
\widetilde{F}_{\mu \nu },
\ee
where $\overline{\theta }= \theta +\textrm{argdet}(M)$, and $M$ is the 
complex, non-hermitian quark mass matrix, arising
from the spontaneous breaking of the electro-weak gauge symmetry.
The real constant $\overline{\theta }$ is an additional 
parameter of the Standard Model, which has to be fixed by experiment.
At the moment, the most constraining bounds on this quantity
come from measurements of the neutron electric dipole moment
 (EDM)~\cite{EDMexp},  which indicate  that  
$\ta~<10^{-9}$~\cite{baluni,witten}.
In this Rapid Communication
we explore some dynamical consequences of the $\theta$-term at the microscopic
level. 
In particular, we identify the mechanisms 
which gives rise to a neutron EDM in the semiclassical limit of QCD, in which
the mixing between 
the degenerate  vacua in (\ref{thetavacuum})
is mediated by instantons. 

Our main result is the discovery that, 
during the tunneling processes,
the $\theta$-term (\ref{Ltheta})
generates an effective repulsion between matter and anti-matter, 
in the neutron. As a consequence, 
quarks and  anti-quarks migrate in opposite directions giving rise to
a finite EDM. 
Our results show that, at least on the semi-classical level,
the EDM arises from the local separation
of positive and negative baryonic charges in the neutron. 
It does not follow from the displacement of the positive and
negative electric charge carried by  
the {\it valence} quarks, as one would intuitively expect in a naive
non-relativistic quark model picture.
\begin{figure*}
\includegraphics[scale=0.33,clip=]{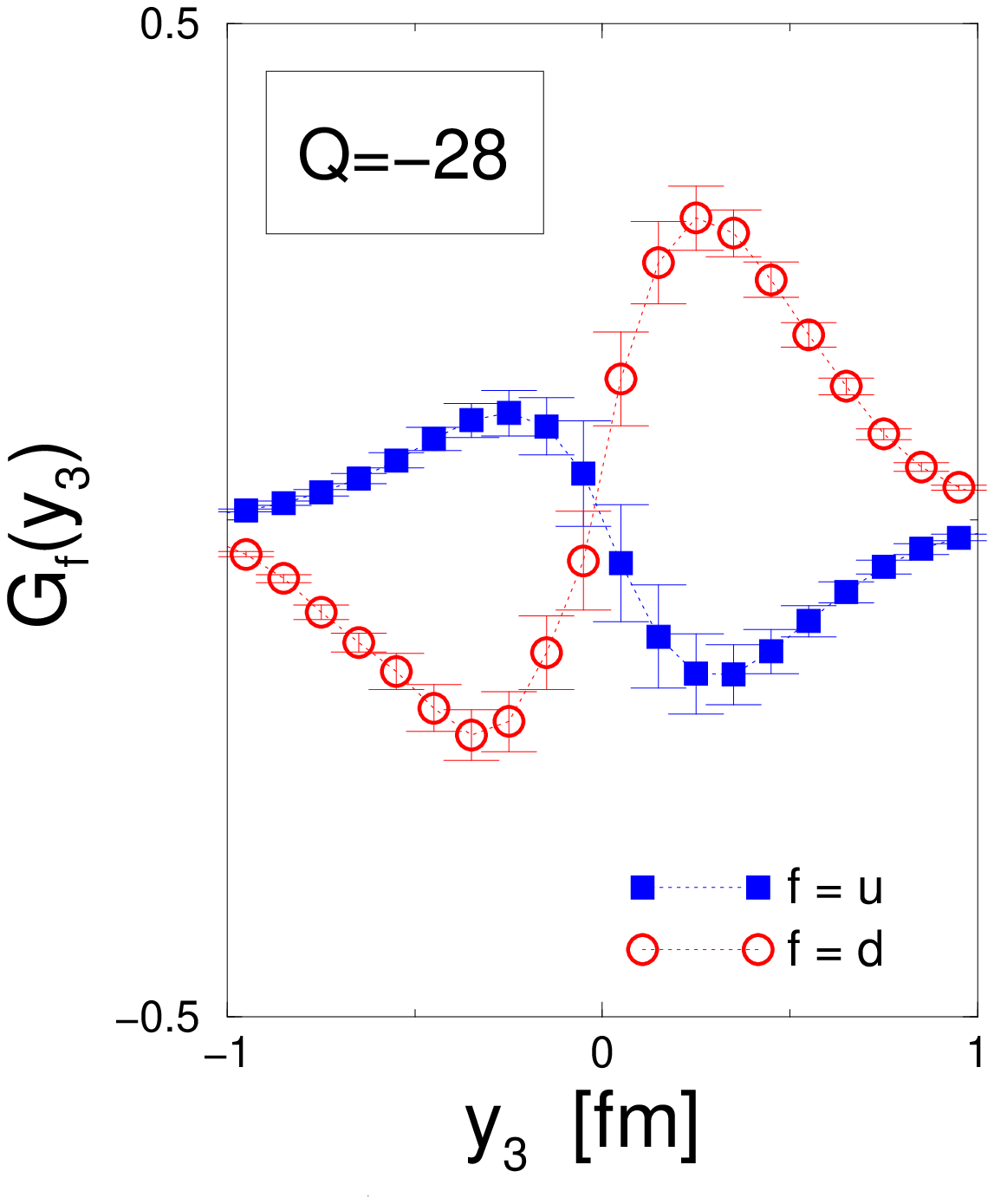}
\hspace{1cm}
\includegraphics[scale=0.33,clip=]{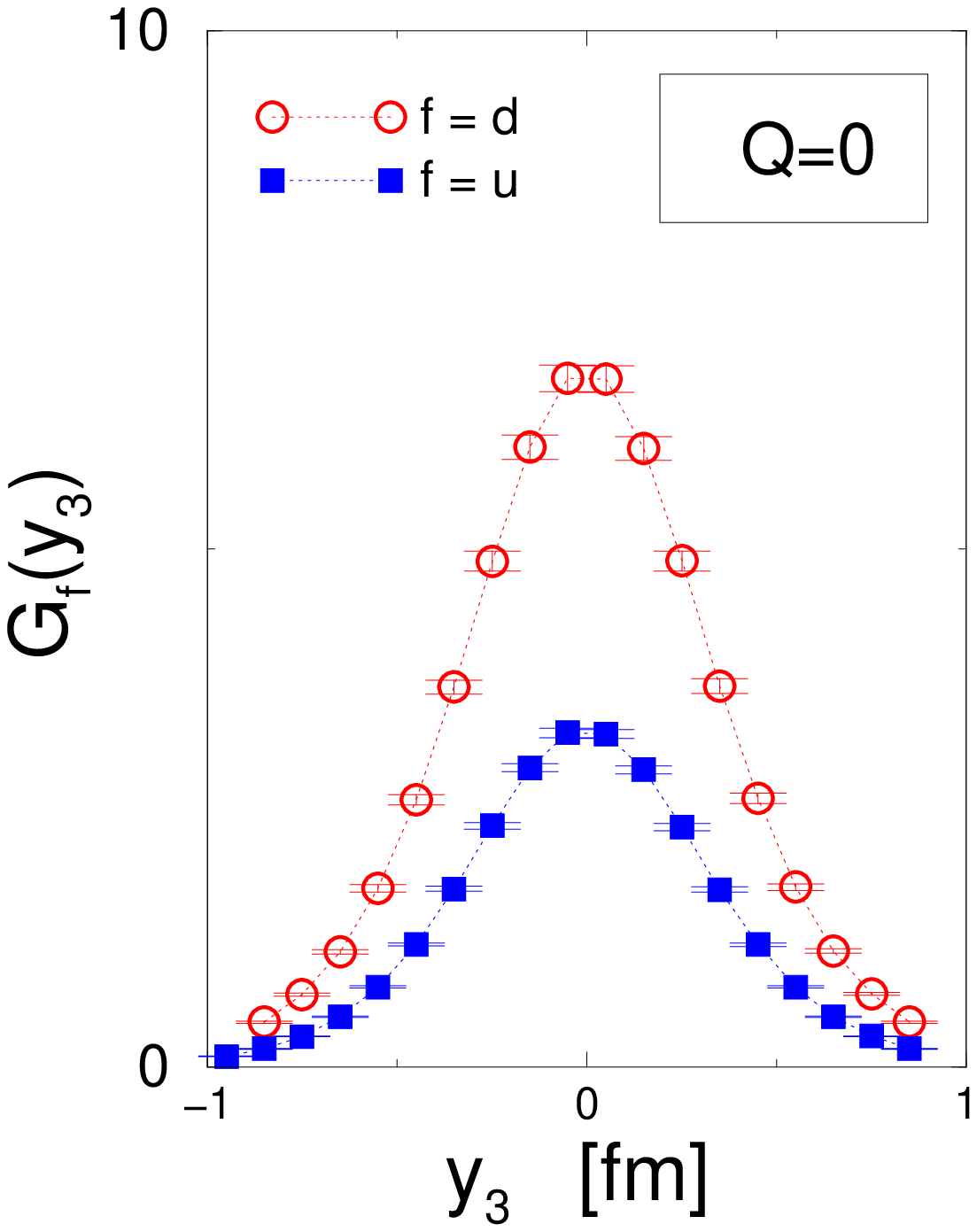}
\hspace{1cm}
\includegraphics[scale=0.33,clip=]{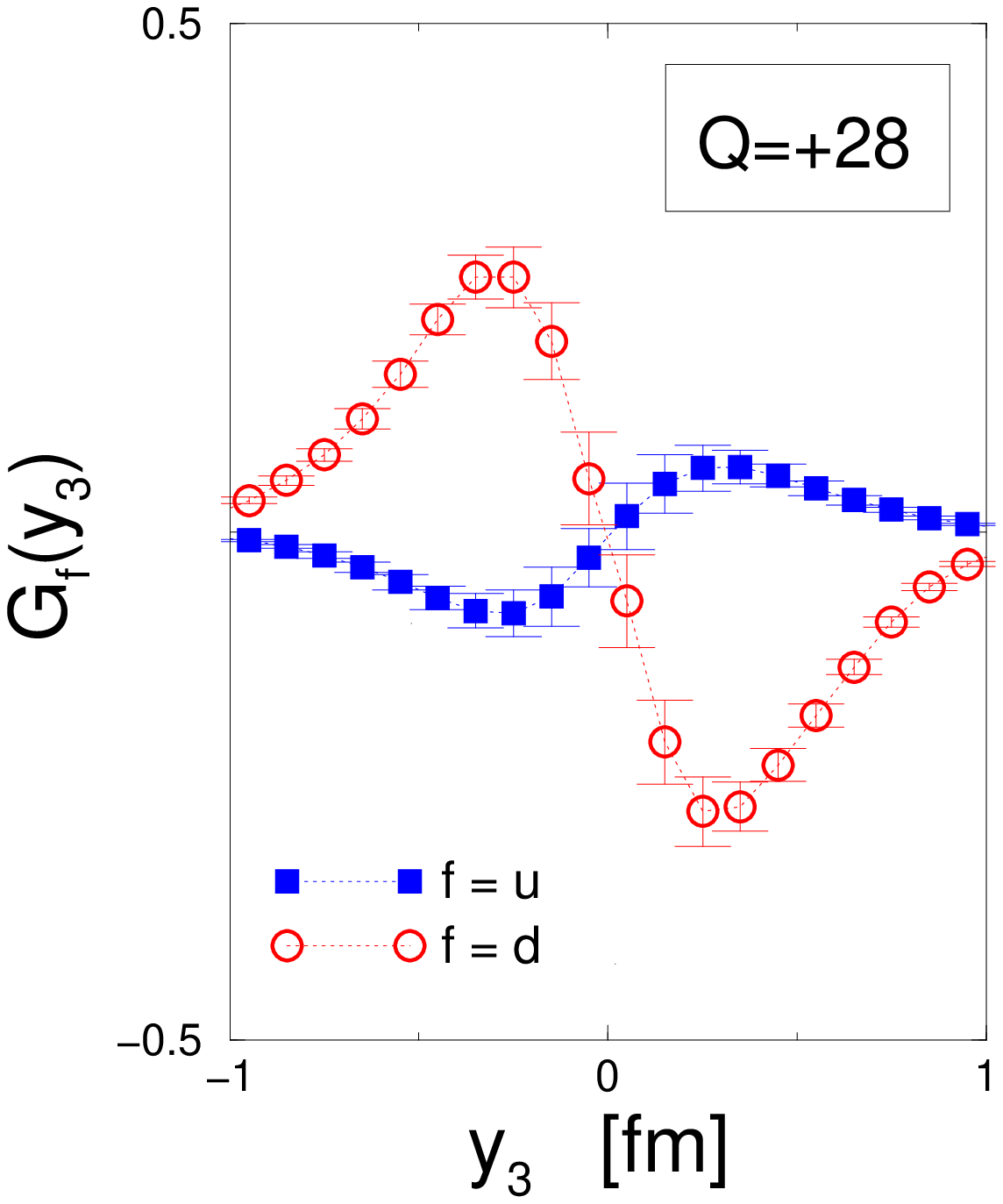}\\
\hspace{1.2cm}\Large{(A)}\hspace{4.5cm} \Large{(B)}\hspace{4.5cm} \Large{(C)}
\caption{The baryon number density
correlators for $u$ and $d$ quarks, ~Eq. (\ref{G3f}),  in units of 
$\ta\times~10^{-5}~\textrm{fm}^{-12}$,  calculated in
canonical ensembles with topological charge 
$Q=-28$, $Q=0$, $Q= +28$.} 
 \label{baryon28}
\end{figure*}
These results are obtained comparing the density distributions of the 
baryonic and electric charges.
Our starting point is the neutron baryon number density correlator,  
defined as 
$G_{B}(\mathbf{y},\tau)=1/3~G_{d}(\mathbf{y},\tau)+1/3~G_{u}(\mathbf{y},\tau)$,
where
\be
\label{G3f} 
G_f(\mathbf{y},\tau)=\langle 0|\;\textrm{Tr}[J_N(\mathbf{0},2\tau )\,
J_4^{f}(\mathbf{y},\tau )\,\overline{J}_N(\mathbf{0},0)\Gamma _3]|0\rangle.
\ee

In (\ref{G3f}) $\tau$ is the Euclidean time, 
$J_\mu^{f}(x)=\overline{q}_f\gamma _\mu\,q_f$, 
$\Gamma_3$ is a diagonal matrix of spinor indices
$\Gamma_3=\textrm{diag}(1,-1,1,-1)$ and
$J_N(x)=\epsilon _{a\,b\,c}(u_a^{T}C\gamma_5\,d_b)\,d_c$
is an interpolating field, which excites states with
the quantum numbers of the neutron. 
Similarly, we define the electric 
charge density correlator as 
$G_{e/m}(\mathbf{y},\tau)=2/3~G_{u}(\mathbf{y},\tau)
-1/3~G_{d}(\mathbf{y},\tau)$.
The correlation function (\ref{G3f}) measures the probability amplitude
to find a quark of flavor $f$ at the point $\bf{y}$, in a system with 
neutron quantum numbers.

The component of the EDM along the 
direction of the neutron spin
can be extracted from the Green's function:
\be
\label{Sigma}
{\bf \Sigma}(\tau) = \int
d^3 {\bf y}\,y_3\, G_{e/m}({\bf y},\tau),
\ee
which, in the large time $\tau$ limit, reads:
\be
{\bf \Sigma}(\tau)\rightarrow D_z\,\times
\,\int 
\frac{d^3\mathbf{p}}{(2\pi )^3}\,
\,
\frac{-2\,\Lambda_N^2\,(p_1^2+p_2^2)}
{e^{2\,\omega_{{\bf p}}\,\tau }\,(\mathbf{p}^2+M^2)},  
\label{const}
\ee
where $D_z$ is the component of the neutron EDM along the spin direction, $M$
is the neutron mass and
$\Lambda_N$ is the coupling of the neutron to the interpolating 
operator, $\la 0|J_N(0)|N\ra= \Lambda_N\,u_{\bf p}$.

In QCD, a  finite dipole can
only arise from the CP-breaking 
interaction (\ref{Ltheta}). To lowest order in $\ta$, we can
write (in an obvious notation): 
\begin{eqnarray}
G_{f}(\mathbf{y},\tau) \simeq 
\frac{-i\,\overline{\theta }}{32\pi ^2}\,\langle 0| 
\textrm{Tr}~[J_N\;J^{f}_4\,\overline{J}_N\,\Gamma_3]
\int d^4 z\,F\widetilde{F}~|0 \rangle _{\overline{\theta }=0}.~\nonumber \\
\label{EDMtheta}
\end{eqnarray}

In the semi-classical limit, the topological charge is condensed around
instantons and anti-instantons ($Q=N_I- N_A$), 
and the physics of the quantum mixing
of the $\theta$-vacuum can be formulated in terms of an
intuitive pseudo-particle picture.
The path-integral  
can then be computed by
summing over the configurations of a statistical 
grand-canonical ensemble of  pseudo-particles. 
In such an approach,
a {\it quantitative} estimate of the matrix element (\ref{EDMtheta}) 
can only be performed in a model-dependent way, as we do not
know from first principles the
density and size distribution of pseudo-particles in the vacuum. 
This is the starting point of the Instanton Liquid
 Model~(ILM) -~for a review see \cite{shuryakrev}~- which 
has been proved to be very successful in describing the phenomenology of 
the QCD vacuum and of light hadrons.

In the present study, we choose to avoid introducing model-dependent parameters
and we focus on the {\it qualitative} effects
 generated by the dynamical interplay of the
strong CP-breaking interaction (\ref{Ltheta}) with the quantum 
structure of the $\theta$-vacuum, at semi-classical level.
A quantitative, albeit model-dependent, prediction of the neutron EDM in this 
model will be presented in a separate publication~\cite{edm}.

In the instanton vacuum, matrix elements with one insertion of
the topological charge operator can be written as a sum over the
contributions of the different
topological sectors~\cite{dyakonov}:
\begin{equation}
\left\langle \mathcal{O}\,\frac 1{32\pi ^2}\int d^4z\,F_{\mu \nu }\widetilde{%
F}_{\mu \nu }\right\rangle =\sum_Q \mathcal{P}(|Q|)\,Q
\left\langle \mathcal{O}\right\rangle_Q ,  \label{topoweight}
\end{equation}
where $\mathcal{P}(Q)$ denotes the relative occurrence of
configurations with topological charge $Q $ and 
$\la \mathcal O\ra_Q$ is the average performed in a canonical ensemble
with such a total topological charge. 
This equation expresses the fact that
CP-breaking interactions  arise from topologically non-trivial gauge 
configurations. Physically, it implies that these forces are
 triggered only during tunneling between distinct classical vacua.

As long as we are interested in qualitative phenomena, we do not need detailed
knowledge of the positive-definite weight factor $\mathcal{P}(|Q|)$ in 
(\ref{topoweight}).
We shall therefore concentrate on the 
contribution from each topological sector (factor 
$Q\,\la\mathcal{O}\ra_Q$ in 
(\ref{topoweight})). This can be done by evaluating
averages in different canonical ensembles in which the number of instantons
and anti-instantons is fixed.
In order to do so,
we have used the  
Interacting Instanton Liquid Model (IILM), developed in \cite{IILM}. 
We have averaged
over 1000 configurations of an ensemble of 130 
pseudo-particles in a periodic box of volume $V=3.2\times~4~\textrm{fm}^4$.
For example, the contribution of the topological sector $Q=10$ was obtained
inserting 70 intantons and 60 anti-instantons in the box.
For each configuration, the fermionic determinant 
and the quark propagator  have been calculated by diagonalizing 
numerically the Dirac operator (for a detailed discussion of the method
see Section VI in~\cite{shuryakrev}). In topologically non-trivial 
sectors, the quark propagator receives contribution from $N_f |Q|$
{\it exact} zero-modes (semi-classical realization of the Index Theorem).
The ensemble of collective coordinates
of the pseudo-particles was then generated dynamically, through an 
usual accept/reject Metropolis algorithm.
To improve the signal-to-noise ratio, we have used rather large quark masses,
$m_u=m_d=75~\textrm{MeV}$ and $m_s=150~\textrm{MeV}$,
and a rather small Euclidean time, $\tau=0.7~\textrm{fm}$. We note that, 
in the present approach, we do not make use of an 
anomalous Ward identity to relate the topological charge to the
divergence of a gauge-invariantly defined axial current. Hence, the
results of the present EDM calculation do not allow to 
distinguish a spontaneous~\cite{crew1} 
from an anomalous~\cite{tHooft} instanton-induced breaking 
of the axial symmetry (for an example in which the two types of breaking lead
to different predictions, see~\cite{bass}).

\begin{figure*}
\includegraphics[scale=0.28,clip=]{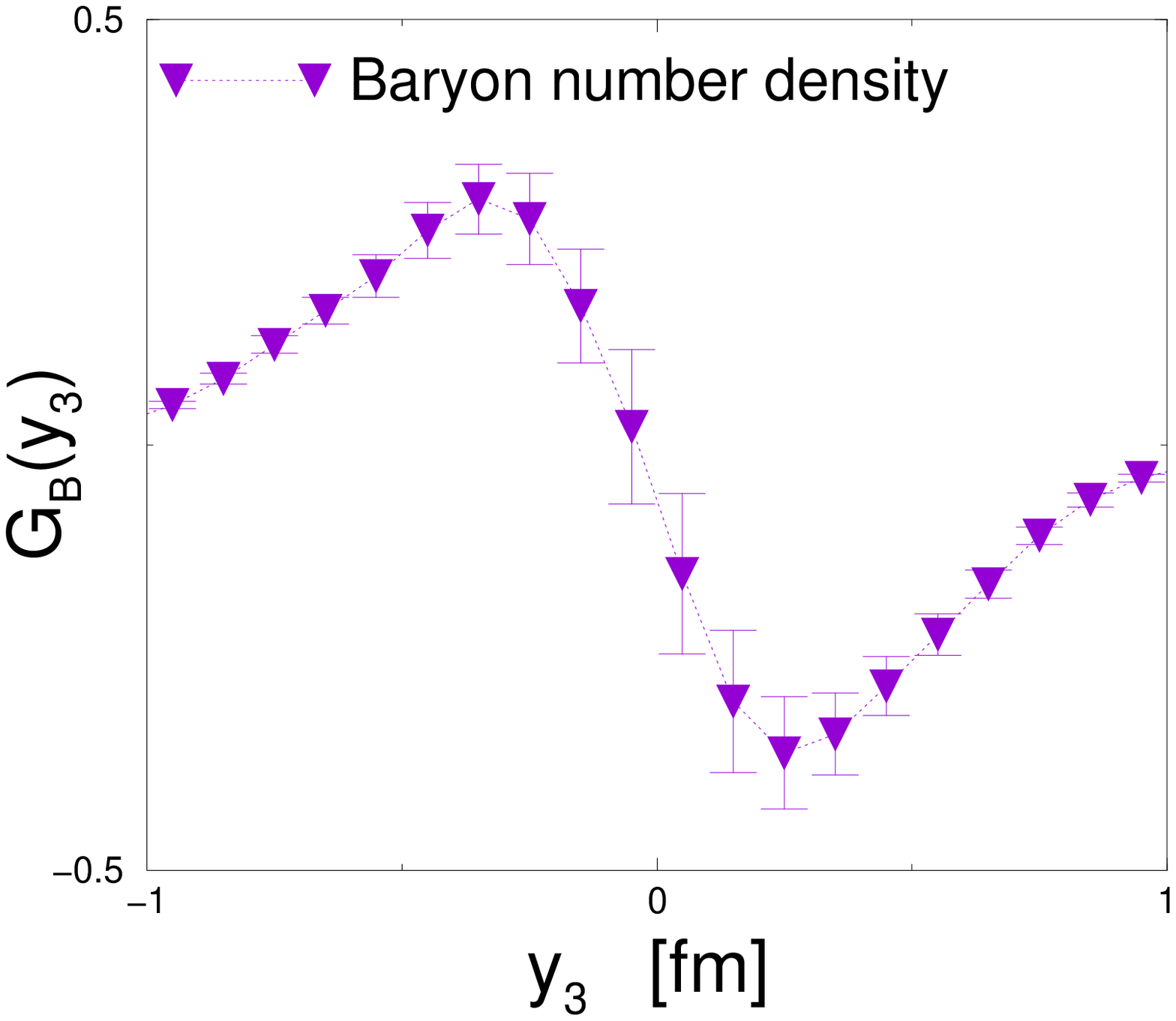}
\hspace{2cm}
\includegraphics[scale=0.28,clip=]{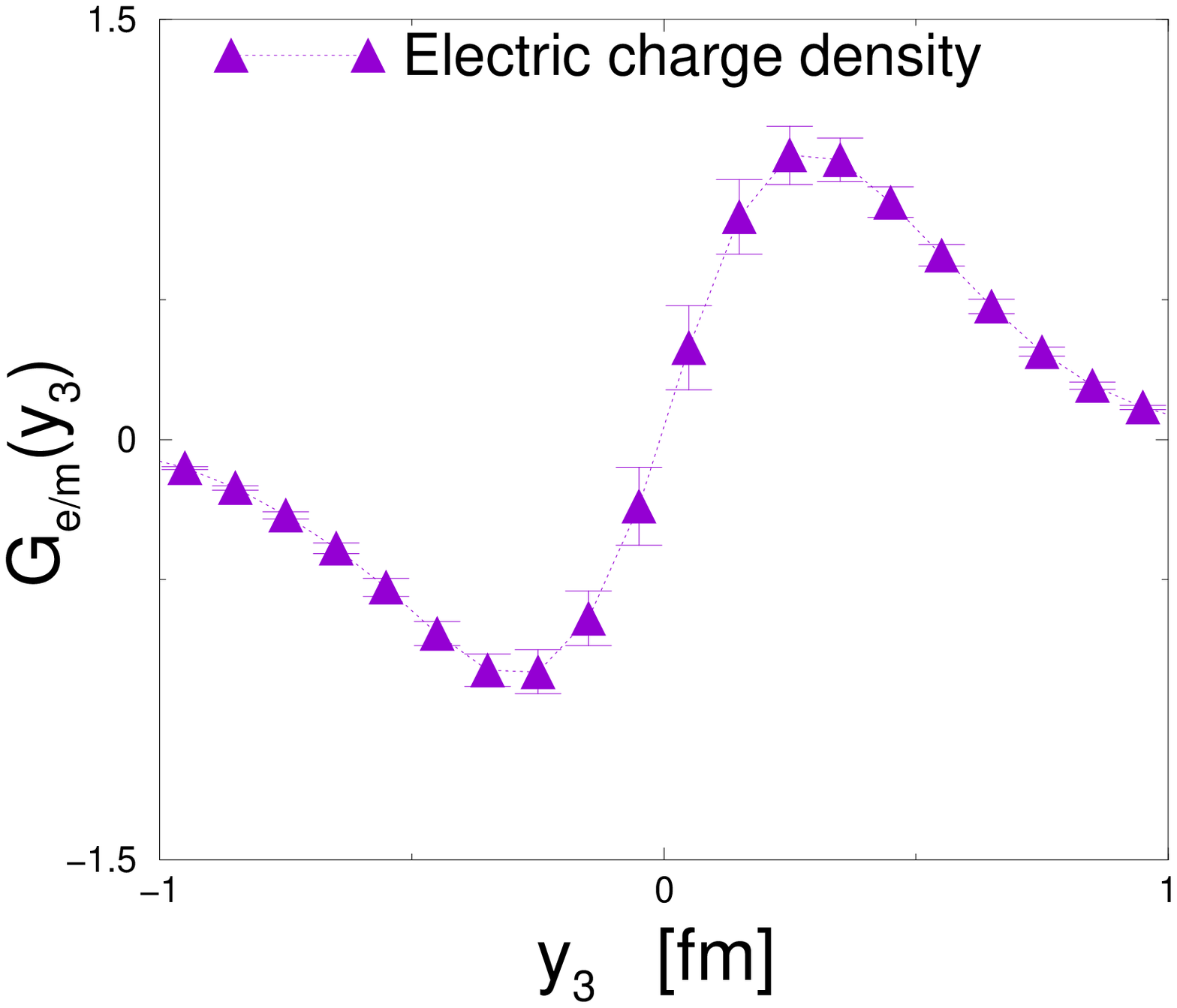}\\
\hspace{1cm}{\Large(A)}\hspace{6cm}{\Large (B)}
\caption{Contribution of the topological 
sectors $Q=-28$ and $Q=+28$ to 
the neutron baryon number density correlator (left panel) and
to the neutron electric charge density correlator (right panel),
in units of $\ta~\times~10^{-5}~\textrm{fm}^{-12}$.
A negative baryon number density for $y_3>0$
denotes a local accumulation of anti-matter
in the positive $y_3$ hemisphere.}
\label{NeutronCharge}
\end{figure*}

In Fig.~\ref{baryon28} we report the results of the  
averages  of the baryon 
density correlators (\ref{G3f}) 
performed in canonical ensembles with 
topological charge $Q=-28$, $Q=0$, $Q=+28$ 
 (~with ${\bf y}$ chosen along the $\hat{z}$ direction~). 
Calculations in ensembles with other positive 
and negative total topological charge have also been performed and
the results were found to follow the same trend.
 Hence, at a qualitative level, Fig.~\ref{baryon28}A (Fig.~\ref{baryon28}C) 
can be regarded as representing {\it all}
terms $\la \mathcal{O}\ra_Q$ in (\ref{topoweight}) with $Q<0$ ($Q>0$). 
(We choose to plot the results of the topological sectors with a
 large topological charge, because they have a better signal-to-noise ratio.) 

These results have several interesting implications.
First of all, we note that the baryon density at zero net topology 
(~Fig.~ \ref{baryon28}B~) is even under parity transformation $y_3\to -y_3$, as
expected.
On the other hand, contributions from the topologically non-trivial
sectors are $P$-odd and lead to a non-vanishing EDM.

The most interesting feature of these results is the separation of positive
and negative baryonic charges in physical systems with neutron quantum numbers.
Fig.~\ref{baryon28}A and Fig.~\ref{baryon28}C imply
that in sectors with non-vanishing  topological charge,
the baryonic charge density of $u$ quarks is positive for $y_3>0$ 
and negative for $y_3<0$. 
Conversely, the baryonic charge density of $d$ quarks is negative 
for $y_3>0$ and positive for $y_3<0$.
Notice that this qualitative effect is the same in all positive
and negative topological sectors, due to the sign carried by the topological
charge factor $Q$  in (\ref{topoweight}).

The physical interpretation of these results is the following.
Any time the quantum QCD ground-state re-arranges itself by tunneling,  
the $\theta$-term (\ref{Ltheta}) generates an 
interaction which effectively 
shifts the  $u$ quarks and the $\bar{d}$ anti-quarks along 
the positive direction of the neutron spin.
At the same time, $\bar{u}$ anti-quarks and $d$ quarks are shifted
toward the opposite direction. 
The net effect is the creation
of an electric current inside the neutron, pointing in the direction
of the spin.

From these correlators we can
construct the contribution to the baryonic number density 
(~Fig.~\ref{NeutronCharge}A~).
We conclude that the tunneling
 produces a local separation of the baryonic charge, with an
accumulation of matter in the 
$y_3<0$ hemisphere 
 and of anti-matter in the $y_3>0$ hemisphere. 
Notice that the contribution to the total
 baryon charge of the neutron coming 
from configurations with $Q\ne0$ vanishes, due to the odd symmetry of the
correlators (\ref{G3f}) under $y_3\to-y_3$ transformations. In other words,  
the neutron {\it total}
 baryon number comes entirely from the topological sector 
with $Q=0$. Hence, the baryon and electric charge asymmetries are associated 
with the sea quarks. 

In Fig.~\ref{NeutronCharge}B we show that the 
separation of the baryonic charge induced by the $\theta$-term 
generates a disentanglement of positive and negative
electric charge in the neutron. This is the microscopic
dynamical mechanism  underlying the EDM formation in QCD, 
at the semi-classical level. 

Notice that in the present discussion we have 
never assumed that the quarks in the 
correlator (\ref{G3f}) form a bound-state.
Indeed, these arguments hold also for 
short-sized Euclidean correlators, in which 
the lowest-lying pole is not isolated from its excitations. 
This means that the dynamical mechanism 
analyzed here does not only concern neutrons, but applies  to all
systems with nucleon quantum numbers. 
In particular, it should be effective also in the its excited states 
and in the proximity of the 
de-confinement  phase transition, where the nucleon is melted into
its partonic components.

We insist on the fact that the outcome 
of our analysis does not depend on the particular
values of the model parameters which define the ILM. 
These results rely only on the working assumption that the quantum 
mixing  of the
QCD vacuum can be described in terms of isolated tunneling events.
On the one hand, there is no a priori guarantee that 
a semi-classical description may
lead to \emph{quantitatively} realistic predictions. 
On the other hand, such an approach contains the correct ingredients
to point out explicitly the connection between the quantum structure 
of the vacuum and the effect of
 CP-breaking interactions. We believe that it is unlikely that 
further quantum corrections 
would completely destroy the qualitative mechanism which
emerges from the present semi-classical analysis.

\begin{figure}
\includegraphics[scale=0.27,clip=]{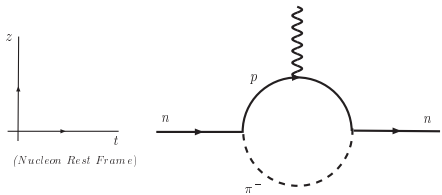}
\caption{Example of a process leading to a positive neutron EDM, 
in the low-energy effective description.}
\label{EDMfig}
\end{figure}

An interesting question to ask is how this microscopic 
QCD dynamics is translated in the language of 
a low-energy effective theory, with hadronic degrees of freedom.
In \cite{witten} Crewther {\it et al.} performed the calculation of the
lowest-order contribution to the EDM in chiral perturbation theory.
In their approach, the EDM arises from the dissociation of the electric charge
of the neutron
associated with the quantum fluctuation, $n\to\pi^-p$ (see Fig.~\ref{EDMfig}). 
In order to produce a finite EDM, 
the neutron must dissociate into charge components, in a way
that breaks the spherical symmetry of the electric charge distribution.
In particular, in order to generate a positive EDM, the virtual 
$\pi^-$ cloud must be mostly localized in the southern hemisphere ($z<0$), 
relative to the direction of the spin of the nucleon.
The same effect is achieved at the microscopic level  by
the non-perturbative instanton-induced effective repulsion
discussed above.
In fact, we have seen that the topological fluctuations drive the 
 $\bar{u}$ and $d$ {\it sea} quarks towards the region $z<0$.
(see Fig.~\ref{baryon28}~C). 

Let us now discuss some phenomenological 
implications of the non-perturbative CP-breaking dynamics discussed 
in this work.
One immediate consequence  is that  $\bar{\theta}>0$ would imply 
an asymmetry in the photon-production of charged pions, with an excess of 
$\pi^-$ produced in the $z<0$ emisphere. This effect is of the same type 
of the asymmetries discussed in~\cite{dima}, 
in the context of relativistic heavy ion collions.

A second phenomenological implication of the present semi-classical 
description  is that the neutron EDM must have
characteristic frequency of
oscillation.
We have seen that the dissociation 
of the electric charge is realized through {\it periodic} currents, 
induced by {\it local} topological  fluctuations in the vacuum. 
When the $Q=0$ 
condition is restored,
the electric charge distribution is relaxed to its 
symmetrical equilibrium state. 
Hence, we expect a natural frequency of oscillations of the EDM, 
of the order of the inverse of the topological screening lenght.
Such a frequency can be estimated in different ways.
For example, in \cite{chimix} we analyzed some combination of 
meson point-to-point 
correlators which relate directly to the amplitude for chirality-flips in
a quark-antiquark state induced by topological interactions.
From a spectral analysis of such a correlation function it was shown
that the information about the initial chiralities of the quark and 
antiquark is exponentially destroyed in time 
with a characteristic decay-constant given by 
$\tau_\chi\sim 1/m_{\eta'}\sim~1/m_{a_0}$ 
(~$a_0$ being the lightest $I=1$, $J^P=0^+$ meson~).
Hence, we can argue that the
characteristic semi-classical frequency of oscillation of the neutron EDM is 
of the order $1/m_{\eta'}$~GeV.

A natural consequence is that, for $\ta\ne~0$, 
we expect a resonance in the neutron Compton scattering cross section,
at center of mass energy 
$\sqrt{s}\simeq M+m_{\eta'}~\simeq~2~{\textrm GeV}$.
The corresponding final photon will be emitted in a state 
such that the total parity will not be conserved in the scattering process.
The observability of such a resonance depends on the magnitude of 
$\ta$. Since the measurements of the EDM indicate that 
this parameter is at least extremely small (if not vanishing), 
it is hard to imagine that any sizable effect could be seen in 
a foreseeable scattering experiment. On the other hand, it would be 
interesting to study what contraints on $\ta$ could be 
set from Compton scattering data. Such a quantitative analysis would
require a model-dependent choice of the instanton size and density 
parameters and is outside the scope of the present work.

Summarizing, we have studied the dynamical mechanism for EDM formation 
in QCD, in the presence of a $\theta$-term.
We have found that the baryon number
carried by the sea quarks in the neutron periodically undergoes
a {\it local} rearrangement. 
This fact can be interpreted as due to a flavor-dependent 
quark-antiquark repulsion,
triggered by  tunneling events in the $\theta$-vacuum.
This is the microscopic origin of the 
breaking the spherical symmetry of the electric
 charge distribution, at the semi-classical level. 
The same mechanism can be effective in excitations of the nucleon and 
in the proximity of the QCD phase transition. 
We have estimated the natural frequency oscillation of the EDM
 to be of the order of the $1/m_{\eta'}$ and argued on possible 
phenomenological implications.

We thank D.~Guadagnoli for his help in the initial stage of the project. 
Clarifying discussions with 
D.~Binosi,  H.~Fritzsch, W.~Metag, L.~Pieri, G.~Ripka, T.~Sch\"afer, 
E.V.~Shuryak,  and W.~Weise are also acknowledged. 

{}
\end{document}